\documentclass[sigconf]{acmart}
\acmSubmissionID{papers\_841}

\usepackage{float}
\usepackage{amsmath}
\usepackage{amsfonts}
\usepackage[inline]{enumitem}
\usepackage{subcaption}
\usepackage{multirow}
\usepackage{soul}
\renewcommand{\hl}[1]{#1} % turns off highlighting via the "soul" package
\usepackage{dcolumn}
\newcolumntype{.}{D{.}{.}{-1}}
\usepackage[ruled]{algorithm2e} % For algorithms

\SetAlFnt{\small}
\SetAlCapFnt{\small}
\SetAlCapNameFnt{\small}
\SetAlCapHSkip{0pt}

\copyrightyear{2024}
\acmYear{2024}
\setcopyright{rightsretained}
\acmConference[SIGGRAPH Conference Papers '24]{Special Interest Group on Computer Graphics and Interactive Techniques Conference Conference Papers '24}{July 27-August 1, 2024}{Denver, CO, USA}
\acmBooktitle{Special Interest Group on Computer Graphics and Interactive Techniques Conference Conference Papers '24 (SIGGRAPH Conference Papers '24), July 27-August 1, 2024, Denver, CO, USA}
\acmDOI{10.1145/3641519.3657477}
\acmISBN{979-8-4007-0525-0/24/07}

\newcommand{\Figure}[1]{Figure~\ref{fig:#1}}
\newcommand{\Table}[1]{Table~\ref{tab:#1}}
\newcommand{\Eq}[1]{Eq.~\ref{eq:#1}}

\newcommand{\Sec}[1]{Sec.~\ref{sec:#1}}

\newcommand{\vc}[1]{\mathbf{#1}}
\newcommand{\mat}[1]{\mathbf{#1}}
\newcommand{\R}{\mathbb{R}}
\renewcommand{\c}{\vc{c}}
\def\vv{\vc{v}}

\newcommand{\A}{\mat{A}}
\newcommand{\B}{\mat{B}}

\newcommand{\I}{\mat{I}}

\newcommand{\M}{\mat{M}}
\newcommand{\N}{\mat{N}}
\begin{document}
% Title portion
\title{Compressed Skinning for Facial Blendshapes}

\author{Ladislav Kavan}
\email{lkavan@meta.com}
\orcid{0000-0001-8549-0878}
\affiliation{%
  \institution{Meta}
  \city{Z\"urich}
  \country{Switzerland}
}

\author{John Doublestein}
\email{jdoublestein@meta.com}
\orcid{0009-0007-8517-1704}
\affiliation{%
  \institution{Meta}
  \city{Redmond}
  \country{USA}
}

\author{Martin Prazak}
\email{map@meta.com}
\orcid{0009-0001-7286-8939}
\affiliation{%
  \institution{Meta}
  \city{Z\"urich}
  \country{Switzerland}
}

\author{Matthew Cioffi}
\email{mail@mattcioffi.com}
\orcid{0009-0008-2199-8224}
\affiliation{%
  \institution{Meta}
  \city{Londonderry}
  \country{USA}
}

\author{Doug Roble}
\email{droble@acm.org}
\orcid{0009-0004-3415-4283}
\affiliation{%
  \institution{Meta}
  \city{Sausalito}
  \country{USA}
}

\begin{abstract}
   We present a new method to bake classical facial animation blendshapes into a fast linear blend skinning representation. Previous work explored skinning decomposition methods that approximate general animated meshes using a dense set of bone transformations; these optimizers typically alternate between optimizing for the bone transformations and the skinning weights. We depart from this alternating scheme and propose a new approach based on proximal algorithms, which effectively means adding a projection step to the popular Adam optimizer. This approach is very flexible and allows us to quickly experiment with various \hl{additional constraints and/or loss functions. Specifically, we depart from the classical skinning paradigms and restrict the transformation coefficients to contain only about 10\% non-zeros, while achieving similar accuracy and visual quality as the state-of-the-art.} The sparse storage enables our method to deliver significant savings in terms of both memory and run-time speed. We include a compact implementation of our new skinning decomposition method in PyTorch, which is easy to experiment with and modify to related problems.
\end{abstract}

% ttph://dl.acm.org/ccs.cfm




\begin{teaserfigure}
    \centering
    \includegraphics[trim={0cm 2in 0mm 0mm},clip]{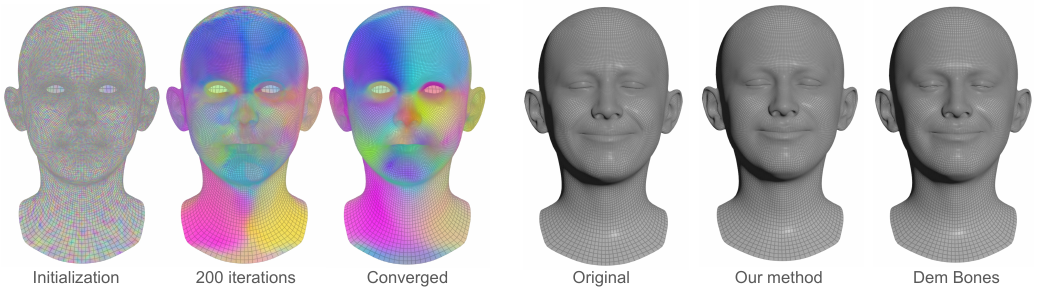}
    \caption{Starting from a random initialization of skinning weights, our method converges to a linear blend skinning approximation of input blendshapes. The accuracy and visual quality of our results is comparable to the state-of-the-art (Dem Bones), but introduces significantly smaller run-time overheads. \hl{In the figure, each color visualizes the influence of one bone, with 40 bones total.}}
    \label{fig:teaser}
\end{teaserfigure}

\maketitle
\section{Introduction}

Many people can access interactive graphics only through inexpensive handheld devices, which is problematic for delivering appealing animated 3D characters. While cloud compute is one potential solution, scaling graphics workloads to billions of users would require a prohibitive amount of compute resources. Therefore, on-device execution is the most viable path to deliver interactive 3D experiences to a large user base. Artists working with facial animation and rigging usually seek maximum creative control and flexibility. This often leads to the use of a large number of blendshapes. High-end film rigs use over 1000 blendshapes, and even less detailed stylized characters still require about 200-300 blendshapes, especially due to corrective shapes needed to overcome the limitations of linear blending. 
%The memory requirements of storing all of these blendshapes are typically not the main limiting factor; a bigger concern is the time to access the blendshape data in run-time, because memory reads are expensive, especially on mobile platforms.

Many data compression algorithms have been studied in the past, but real-time animation requires a decompression method that is fast and integrates well with a GPU that performs the final rendering. Linear blend skinning decomposition methods \cite{james2005skinning} have been proposed to solve this problem for general animated meshes and they also work well for facial blendshapes. Linear blend skinning decomposition \hl{departs from the traditional hierarchical skeleton structures and instead} approximates an input data matrix \hl{$\A \in \R^{3S \times N}$ as a product of two factors: $\A \approx \B \mat{C}$, where $\B \in \R^{3S \times 4P}, \mat{C} \in \R^{4P \times N}$. The input matrix $\A$ contains the x, y, z coordinates of the $S$ input shapes with $N$ vertices; the $P$ is the number of proxy-bones.} The matrices $\B$ and $\mat{C}$ have special structure: $\B$ is typically dense and stacks the skinning transformations; $\mat{C}$ is typically sparse and combines the skinning weights with the rest pose. The problem of ``skinning decomposition'' can be defined as finding the factors $\B$ and $\mat{C}$ for a given $\A$. The state-of-the-art method for solving this problem is implemented in the open source `Dem Bones'' library \cite{le2012smooth,dembones} which has been integrated in Maya, Houdini and many other tools. Linear blend skinning is a widely recognized standard in interactive graphics. By adhering to this standard, we can take advantage of highly optimized shader implementations, file formats, and tools that are already in place.
%Therefore, the linear blend skinning decomposition became a popular method in the industry, also due to the excellent open source ``Dem Bones'' library \cite{dembones} which has been integrated in Maya, Houdini and many other tools. 

In Dem Bones, as well as in most publications on this problem, the user has to select a fixed number of proxy-bones. These proxy-bones do not have any anatomical meaning, but they correspond to individual skinning transformations; each transformation drives part of the mesh. With our rigs, we found that as few as 40 proxy-bones lead to acceptable quality with Dem Bones. However, even with a modest set of 300 blendshapes, this introduces significant run-time overheads, because for each of the shapes we need to store all of the proxy-bone transformations: $300 \times 40 = 12000$ (with each individual transformation represented either via a $3 \times 4$ matrix or a [quaternion, translation] pair). These transformations have to be read and blended at run-time for each character, every frame.
We can do better if we revisit the problem of skinning decomposition from scratch.
%One important feature of classical skinning is sparsity of the $\mat{C}$ matrix, which corresponds to the fact that each proxy-bone can affect only a limited number of vertices. 
Previous work explored solvers based on alternating between optimization for the skinning weights and the transformations (cyclic coordinate descent). More and more efficient and general solvers were explored during the 2005 - 2015 decade and open sourced in 2019 in the Dem Bones library. The need to minimize the run-time compute requirements lead us to explore an entirely different optimization strategy, motivated by the success of first-order optimization methods used in deep learning. This is not trivial, because typical deep learning optimizers assume unconstrained optimization and typically handle bounds via sigmoid or softmax functions. This may be sufficient for non-negativity and affinity of the skinning weights, but the hard constraint of spatial sparsity is more difficult. Inspired by proximal algorithms, well studied in the convex setting \cite{parikh2014proximal}, we append a constraint-projection step to the popular Adam optimizer \cite{kingma2014adam}; this is a key ingredient of our proposed method.

We are solving exactly the same optimization objective (loss) as previous skinning decomposition methods, including Dem Bones. However, this objective is highly non-convex and thus different optimization methods converge to different solutions. Our approach consistently converges to solutions with lower errors than the Dem Bones solver. We can even lower the number of proxy-bones to 20 and still obtain similar accuracy as Dem Bones, but we needed more significant savings. Our key finding to report in this paper is that we can introduce sparsity constraints also on the transformations, i.e., zero-out about 90\% of the coefficients of $\B$ and still match the accuracy of the Dem Bone results (this is in addition to the sparsity of $\mat{C}$, i.e., both $\B$ and $\mat{C}$ are now sparse). \hl{This additional sparsification is straightforward to implement as part of the constraint-projection step.}
Even when accounting for the overhead of sparse data structures, this represents significant savings at runtime. To our knowledge, this additional ``skinning transformation sparsification'' has not been explored before and we propose to name it ``compressed skinning'' to distinguish it from previous skinning decomposition methods which compute dense $\B$.

Our strategy of leveraging deep learning optimizers allows for a compact implementation in PyTorch \cite{paszke2019pytorch} with automatic backward differentiation and trivial deployment to CUDA. There are also additional benefits: previous skinning optimization methods start with a sophisticated initial guess computed with spectral or k-means clustering. With our approach, we simply use the Gaussian noise to initialize all of our variables (\Figure{teaser}) -- more complicated initializations were not necessary and can introduce bias or even lead to worse results (e.g. in the case of symmetries preserved by the optimizer). Another advantage of our approach is its flexibility, e.g., changing the loss function is just a simple one-line code modification. We leveraged this flexibility and tried improving the accuracy of our fits by changing the error metric from the classical $L^2$ norm to an $L^p$ norm with higher $p$. When we changed the norm to $L^{12}$ and \hl{increased the number of influences to 32, the maximal error dropped by more than 50 times compared to Dem Bones, allowing us to recover even fine wrinkles in the original blendshapes, at the cost of more non-zeros weights in matrix $\mat{C}$.} However, our main focus is not improvements of quality but rather run-time efficiency. 

Another contribution is a precise formulation of converting the ``blend-weights'' (i.e. the time-varying blendshape coefficients) to linear blend skinning transformations. This problem was not considered in previous skinning decomposition methods \cite{james2005skinning,le2012smooth} which were more general and not focused on blendshape animation. This conversion is not hard but also it is not trivial because 1) skinning handles the rest pose differently than blendshapes and 2) linear blending of transformations is correct only when using a carefully chosen subset of spatial transformations (specifically, we use linearized rotations). 
%Even though our first-order optimization approach requires longer pre-processing time than Dem Bones (on the order of minutes on a single GPU), we believe this is practically irrelevant compared to the benefits of reducing the compute footprint on target devices, helping us to deliver delightful 3D characters to a large population of users.
\hl{Our first-order optimization approach requires longer pre-processing time than Dem Bones (on the order of minutes on a single GPU), but reduces the computation footprint on target devices.}

Our main technical contributions are:
\begin{itemize}
    \item New, first-order optimization method that solves skinning decomposition better than previous methods
    \item New skinning decomposition constraint: sparsity of transformations (``compressed skinning'')
    \item Explicit formulas to convert facial animation blend-weights into skinning transformations
\end{itemize}

\hl{However, our method is more complex than classical linear blend skinning and drifts away from established practices, in particular:}
\begin{itemize}
    \item \hl{Our method needs an extra sparse matrix-vector multiplication on the CPU to compute skinning transformations.}
    \item \hl{The GPU skinning shaders need to support arbitrary transformation matrices (rigid transformations are not sufficient).}
\end{itemize}

\section{Background and Related Work}

Traditional facial blendshape models are typically rooted in Facial Action Coding System \cite{ekman1978facial} or its variants \cite{lewis2014practice}, though
more recent research aims at addressing their limitations: complex controls and large number of blendshapes needed for high fidelity \cite{kim2021optimizing,choi2022animatomy}.
If training data or facial rigs are available, it is possible to derive more efficient models using \hl{classical statistical methods \mbox{\cite{meyer2007key}}} or deep learning \cite{bailey2020fast,chandran2022facial}. Our approach builds upon deep learning optimizers (Adam), but utilizes only the classical linear blend skinning model -- no neural networks.

The problem of skinning decomposition was introduced by \cite{james2005skinning} and considered general animation sequences, not just facial animation, but also e.g. cloth animation or stylized full-body characters. Subsequent research led to better approximations of the unknown global optimum \cite{kavan2010fast}, studied the benefits of adding bone rigidity constraints \cite{kavan2007skinning,le2012smooth} and further compression of the skinning weights via two-layer skinning \cite{le2013two}. All of the these methods find a cloud of unstructured proxy-bone
transformations. An alternative approach aims to fit input animations with hierarchical skeletons \cite{schaefer2007example,hasler2010learning,bharaj2012automatically,moutafidou2023deep}
which offers better editability and compatibility with tools such as Maya with well established support for hierarchical skeletons. Even though hierarchical skeletons are most commonly used for full-body animation, they can be also useful in facial animation, e.g. for the jaw and the eyes \cite{tianye2017flame}. However, high quality jaw opening can be also achieved without any skeleton using ``intermediate'' blendshapes \cite{lewis2014practice}. \cite{seo2011compression} studied blendshape compression via hierarchical semi-separable matrices, which finds permutations of the blendshape matrix and a tree structure eliminating blocks of zeros. This leads to a compact representation, but their method is closed-source and the decompression algorithm is much more complex than skinning. \hl{More recently, GPU-specific methods to speed up blendshape evaluation using compute shaders were proposed \mbox{\cite{costigan2016improving}}.}
% The open source and linear-blend-skinning-compatible Dem Bones \cite{dembones} became more popular.

Beyond computer graphics, methods such as Lasso \cite{tibshirani1996regression}, non-negative matrix factorization \cite{lee1999learning,hoyer2004non} and overcomplete dictionary learning \cite{aharon2006} discover sparse structures in input data. Compressed sensing \cite{donoho2006compressed} solves underdetermined linear systems by assuming sparsity. The common theme in these
methods are sparsity-inducing norms. The idea of combining the Adam optimizer with a proximal operator has been studied for the source separation problems in astronomical imaging \cite{melchior2019proximal}. Each domain (structured data / images / meshes) has its own specifics. In the context of animated polygon meshes with constant connectivity, a scenario similar to ours, the $L^1$ norm has been utilized to transform an input mesh animation into a blendshape representation \cite{neumann2013sparse}. However, our approach differs in that we take blendshapes as input and convert them into skinning. In other words, our method could potentially follow \cite{neumann2013sparse} or any other method that generates blendshapes.

% , which is characterized by sparsity and locality (in contrast to the global PCA basis \cite{alexa2000representing}). 

% The $L^1$ approach has been also studied in our context of animated polygon meshes \cite{neumann2013sparse}. This well-known
% SPLOCS method converts an input animation to a blendshape representation, which is characterized by sparsity and locality (in contrast to the global PCA basis \cite{alexa2000representing}). 
% Instead, our method assumes sparse and localized blendshapes as input and converts them to skinning.

\subsection{Blendshape Facial Models}\label{sec:BFM}

In this section we briefly recap standard methods for facial blendshape modeling and establish notation for future sections. The classical blendshape model \cite{lewis2014practice} uses the following ``delta'' formulation centered at the rest-pose (neutral facial expression):
\begin{equation}\label{eq:delta-blending}
    \hat{\vv}_{0, i} + \sum_{k=1}^{S} c_k (\hat{\vv}_{k, i} - \hat{\vv}_{0, i})
\end{equation}
\hl{where $\hat{\vv}_{0,i} \in \R^3$ is $i$-th rest-pose vertex, $\hat{\vv}_{k, i} \in \R^3$ is $i$-th vertex in blendshape number} $k$ and $c_k$ are blending coefficients. With all coefficients $c_k$ set to zero, we obtain the rest-pose, typically representing a neutral facial expression. In a basic linear blendshape model, all $c_k$ are directly driven by the animator. However, the well-known limitations of linear blending lead to dissatisfying results for some combinations of $c_k$ values. To avoid this problem, artists typically do not modify the $c_k$ coefficients directly, but create a function known as a ``rig'' that takes as input a reduced, animator-friendly set of control values and outputs all of the $c_k$ coefficients. These controls are typically FACS-based \cite{ekman1978facial} and low-dimensional (about 40 to 80 controls), while the $c_k$ coefficients are more numerous (200 to 300, or more in film-quality models). The rig function is typically made from elementary non-linear blocks supporting intermediate and combination (or ``corrective'') shapes. The details are not important in this paper but are well explained in the literature \cite{lewis2014practice}. 

\subsection{Linear Blend Skinning}\label{sec:LBS}

Linear blend skinning has been originally developed for smooth deformations of articulated meshes \cite{magnenat1988joint}, but it is a general deformation model \cite{jacobson2011bounded}:
\begin{equation}\label{eq:LBS}
    \sum_{j=1}^P w_{i, j} \M_{j} \vv_{0,i}
\end{equation}
\hl{where $\vv_{0,i} \in \R^4$ is $\hat{\vv}_{0,i}$ but with the additional coordinate set to 1;}
$\M_{j} \in \R^{3 \times 4}$ are affine transformation matrices, corresponding to either skeletal bones in full-body animation or virtual proxy-bones \cite{james2005skinning}; $w_{i, j}$ are skinning weights satisfying the following constraints; \textit{non-negativity}: $w_{i,j} \geq 0$, 
\textit{partition of unity}: $\sum_j w_{i, j} = 1$ and \textit{spatial sparsity}: the number of nonzero $w_{i,j}$ for each $i$ is bounded by a constant $K$. \hl{\mbox{\Eq{LBS}} transforms only vertex positions; normals are often approximated but an accurate and efficient method exists \mbox{\cite{tarini2014accurate}}.}

The transformations $\M_j$ are the only time-varying parameters in an animation and thus play the same role as the $c_k$ coefficients in blendshape models (\Eq{delta-blending}). To convert a blendshape model to a skinned one, we will need a way to convert $c_k$ into $\M_j$. To our knowledge, this conversion method has not been presented in the literature; skinning decomposition papers \cite{james2005skinning,le2012smooth} assume general animated meshes and provide only playback functionality -- they do not discuss the specifics of facial models and the combination of blendshapes represented by skinning transformations. 
%Let's dive into this in the next section.

\begin{figure*}
    \centering
    \includegraphics[trim={0cm 2.3in 0mm 0mm},clip]{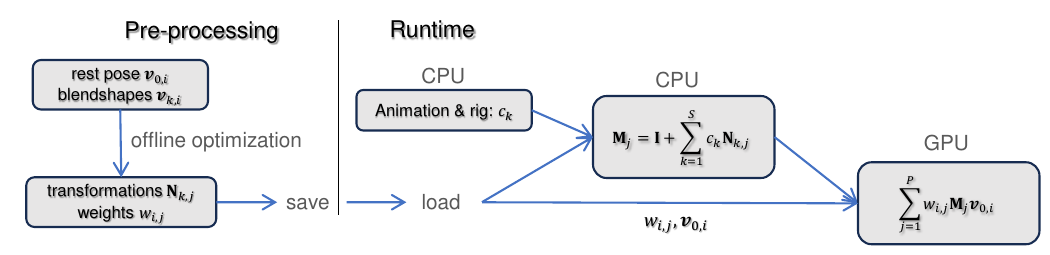}
    \caption{\hl{The skinning decomposition is pre-computed offline (left). On the end-user device, we first load the pre-computed $w_{i,j}$ and $\N_{k,j}$. Then, for each animation frame (runtime, right), we obtain $\c_k$ from the rig and compute $\M_j$. The skinning transformations $\M_j$ along with the rest-pose $\vv_{0, i}$ and weights $w_{i,j}$ are passed to linear blend skinning module running on the GPU.}}
    \label{fig:workflow}
\end{figure*}

\section{Framework}\label{sec:F}

We start by writing down the formula for converting $c_k$ into $\M_j$. This will also help clarify the framework of our approach and explain the setup for our new \textit{compressed skinning decomposition} algorithm (\Sec{CSD}).

We introduced the problem of \textit{skinning decomposition} via a succinct abstraction using matrices $\A \in \R^{3S \times N}, \B \in \R^{3S \times 4P}$ and $\mat{C} \in \R^{4P \times N}$. \hl{The matrix $\A$ contains \mbox{$\hat{\vv}_{k, i} - \hat{\vv}_{0, i}$} and the matrices \mbox{$\B, \mat{C}$} have the following structure:}
\begin{equation}
\B =
\begin{bmatrix}
\hat{\N}_{1,1}    & \dots & \hat{\N}_{1,P} \\
\dots       & \dots & \dots  \\
\hat{\N}_{S,1}    & \dots & \hat{\N}_{S,P} \\
\end{bmatrix},\ 
\mat{C} =
\begin{bmatrix}
w_{1,1} \vv_{0,1}   & \dots & w_{N,1} \vv_{0,N} \\
\dots       & \dots & \dots  \\
w_{1,P} \vv_{0,1}   & \dots & w_{N,P} \vv_{0,N} \\
\end{bmatrix}
\end{equation}
\hl{The matrices $\hat{\N}_{k,j} \in \R^{3 \times 4}$ are defined as $\hat{\N}_{k,j} = \I + \N_{k,j}$, where $\I \in \R^{3 \times 4}$ is a $3 \times 3$ identity matrix extended with a column of zeros. With this notation, the formula $\B \mat{C} = \A$ is equivalent to:}
\begin{equation}\label{eq:N}
    \sum_{j=1}^P w_{i, j} (\I + \N_{k, j}) \vv_{0, i} = \hat{\vv}_{k, i}
\end{equation}
\hl{Due to the fact that $\sum_j w_{i,j} \I {\vv}_{0,i} = \hat{\vv}_{0,i}$ which follows from the partition of unity of the skinning weights,
\mbox{\Eq{N}} is also equivalent to:}
\begin{equation}\label{eq:Ndelta}
\sum_{j=1}^P w_{i, j} \N_{k, j} \vv_{0, i} = \hat{\vv}_{k, i} - \hat{\vv}_{0, i}    
\end{equation}
which shows that \hl{the transformations $\N_{k,1}, \dots, \N_{k,P}$ represent exactly the $k$-th delta-shape $\hat{\vv}_{k,i} - \hat{\vv}_{0,i}$ (\mbox{\Eq{delta-blending}}). Intuitively, the subtraction of the identities $\I$ corresponds to subtraction of the rest pose.} This is related to the fact that linear blend skinning returns the rest-pose for \textit{identity} transformations, while \Eq{delta-blending} returns the rest-pose for \textit{zero} blendweights $c_k$.

Let us now assume that we have solved for $w_{i,j}$ and $\N_{k, j}$ that satisfy \Eq{Ndelta} and all of the skinning weight constraints (this will be discussed in more detail in \Sec{CSD}). Now we just need a formula to compute the final skinning transformations. \hl{This formula can be derived by algebraic manipulations of \mbox{\Eq{delta-blending}}:}
\begin{equation}\label{eq:equiv_proof}
\begin{split}
    \hat{\vv}_{0, i} + \sum_{k=1}^{S} c_k (\hat{\vv}_{k, i} - \hat{\vv}_{0, i}) = \hat{\vv}_{0,i} + \sum_{k=1}^S c_k \sum_{j=1}^P w_{i,j} \N_{k,j} \vv_{0, i} \\
    = \sum_{j=1}^P w_{i, j} \left(\I + \sum_{k=1}^S c_k \N_{k, j} \right) \vv_{0,i}
%    \sum_j w_{i, j} \left(\I + \sum_k c_k \N_{k, j} \right) \vv_{0,i} &= \vv_{0,i} + \sum_k c_k \sum_j w_{i,j} \N_{k,j} \vv_{0, i} \\
%     &= \vv_{0, i} + \sum_k c_k (\vv_{k, i} - \vv_{0, i})
\end{split}
\end{equation}
\hl{Therefore, if we define the final skinning transformations $\M_j$ as:}
\begin{equation}\label{eq:M}
    \M_j = \I + \sum_{k=1}^S c_k \N_{k, j}
\end{equation}
\hl{for $j = 1, \dots, P$, the linear blend skinning formula (\mbox{\Eq{LBS}}) produces the exact same result as the blending of the blendshapes, as shown in \mbox{\Eq{equiv_proof}}. 

The flowchart of our method is in \mbox{\Figure{workflow}}. A limitation of our method compared to a standard skinning implementation is the need to evaluate \mbox{\Eq{M}} on the CPU. In the case of dense $\N_{k,j}$, as with Dem Bones implementation, this is a dense matrix-vector product. We propose sparse $\N_{k,j}$ in which case this becomes a sparse matrix-vector product. In either case, the resulting $\M_j$ are dense and are passed to a standard skinning shader. In typical character animation systems, facial animation is usually composited with a head (or neck) transformations and full-body skinning. This can be accomplished by multiplying our $\M_j$ with the head transformation on the CPU, as in standard hierarchical skeletal animation.}

%On the end-user device, we first load the pre-computed $w_{i,j}$ and $\N_{k,j}$. Then, for each animation frame, we use the rig function to compute $\c_k$ and \Eq{M} to compute $\M_j$. We pass these transformations $\M_j$ along with the rest-pose $\vv_{0, i}$ and weights $w_{i,j}$ to a linear blend skinning module, which is standard and well optimized in most game engines and similar systems. This skinning module evaluates \Eq{LBS}, typically in a vertex shader on a GPU. 

% Due to our setup, the resulting deformed (``skinned'') vertex positions are equivalent to \Eq{delta-blending}, but without using the original blendshapes $\vv_{k,i}$. This equivalence can be verified by simple algebra, plugging \Eq{M} into \Eq{LBS} and using the identity from \Eq{Ndelta}:
% %
% \begin{equation}
% \begin{split}
%     \sum_j w_{i, j} \left(\I + \sum_k c_k \N_{k, j} \right) \vv_{0,i} &= \vv_{0,i} + \sum_k c_k \sum_j w_{i,j} \N_{k,j} \vv_{0, i} \\
%      &= \vv_{0, i} + \sum_k c_k (\vv_{k, i} - \vv_{0, i})
% \end{split}
% \end{equation}
%
In summary, we have converted the delta-blendshape model (\Eq{delta-blending}) to linear blend skinning. This conversion is exact only if \Eq{Ndelta} is satisfied exactly; in practice there will be some errors, but we will minimize them to ensure a visually pleasing approximation of the original blendshape model. Our key contribution (and the rationale behind the name ``Compressed Skinning'') is that most of our final $\N_{k,j}$ transformations will be zero, enabling us to realize significant savings compared to dense methods such as Dem Bones.

\subsection{Transformation representation}

The most general choice for $\N_{k,j}$ are general $\R^{3 \times 4}$ matrices, but some authors proposed restricting the $3 \times 3$ components to rotations \cite{le2012smooth}. This works well for general animation sequences which are \textit{not} driven by delta-blending, but our approach relies on the linear blending in \Eq{M}. Even if we constrained $\N_{k,j}$ to rotations, the resulting $\M_j$ will \textit{not} be rotations due to \Eq{M}. 
%Even though intrinsic blending of rotations e.g. via quaternions is often better than linear blending \cite{shoemake1985animating}, this is not the case here.
Therefore, in our final method we chose to use the following class of transformations:
\begin{equation}\label{eq:rt}
\N_{k,j} = 
\begin{pmatrix}
0          & -r_{k,j,3}    & r_{k,j,2}  & t_{k,j,1} \\
r_{k,j,3}  & 0             & -r_{k,j,1} & t_{k,j,2} \\
-r_{k,j,2} & r_{k,j,1}     & 0          & t_{k,j,3} \\
\end{pmatrix}
\vspace{+1mm}
\end{equation}
using only three degrees of freedom for the linearized rotation ($r_{k,j,:}$) \cite{goldstein2002classical} and another three for the translation ($t_{k,j,:}$). This class of transformations is closed under linear blending, i.e., linear combination of transformations of the form of \Eq{rt} produces another transformation of the same form. This means that \Eq{M} can work with the 6-dimensional representation in \Eq{rt} and compute nothing but plain linear blending at run-time. This is faster than e.g. quaternion interpolation of rotations, which requires a projection/normalization step and safeguards to ensure shortest-path interpolation \cite{buss2001spherical}. In our case, linear blending is correct because the original blendshapes were designed to be blended linearly (\Eq{delta-blending}); non-linearities such as jaw opening are handled in the rig, which we treat as a black box. This is an efficient approach to convert blendshape models into linear blend skinning which also lends itself to a well structured code.

\section{Compressed Skinning Decomposition}\label{sec:CSD}

\hl{In this section we discuss the details of our skinning decomposition (\mbox{\Figure{workflow}}: pre-processing).
%In our setting, the data matrix $\A$ contains the blendshape deltas $\vv_{k, i} - \vv_{0, i}$ and the task of skinning decomposition consists of finding factors $\B$ and $\mat{C}$ such that their matrix product is close to the data matrix in some norm. The $\B$ matrix stacks the individual $\N_{k,j} \in \R^{3 \times 4}$ and, similarly, $\mat{C}$ contains $w_{i, j} \vv_{0, i}$. 
This is a non-convex optimization problem:}
%
%The matrix product of $\B$ and $\mat{C}$ can be written out explicitly using indices and the final minimization problem of skinning decomposition can be written as:
%
\begin{equation}\label{eq:loss}
\min_{w_{i,j}, \N_{k,j}} \sum_{i=1}^N \sum_{k=1}^S (E_{i,k})^p \text{, } E_{i,k} = \lvert \vv_{k, i} - \vv_{0, i} - \sum_j w_{i, j} \N_{k, j} \vv_{0, i}  \rvert
%E_{i,k} = \left\Vert \vv_{k, i} - \vv_{0, i} - \sum_j w_{i, j} \N_{k, j} \vv_{0, i}  \right\Vert_p
\end{equation}
where by default we use $p=2$ corresponding to the standard Euclidean norm, in which case the absolute value in the definition of $E_{i,k}$ is moot. 
%where the optimization variables are the skinning weights $w_{i,j}$ and the transformations $\N_{k,j}$. 
%Even though it would be possible to optimize also the rest-pose $\vv_{0,i}$, we found that this does not improve the results significantly and introduces potential issues with rest-pose degeneracies, so we assume that $\vv_{0,i}$ are fixed to the artist-defined rest-pose. 

\Eq{loss} is straightforward to implement in PyTorch, please see the attached code (function \texttt{compBX}). However, an important feature of the skinning decomposition are the following constraints imposed on $w_{i,j}$: non-negativity, partition of unity and spatial sparsity (\Sec{LBS}). The spatial sparsity ensures the sparsity of $\mat{C}$, which is a classical feature of skinning decomposition \cite{james2005skinning}. On the other hand, the sparsification of $\B$ is a new contribution in this paper and we will discuss it in more detail below. To ensure that the skinning approximation is smooth, we also add a Laplacian regularization term, the same as used in previous work \cite{le2014robust} and in the open source Dem Bones implementation. 
%Note that in this equation, the $\vv_{k, :} - \vv_{0, :}$ corresponds to the $k$-th delta-blendshape (\Sec{BFM}), which is the input to the skinning decomposition algorithm, stored in a data matrix.

Our plan is to leverage the well established Adam optimizer, but the problem is that it is an unconstrained optimizer. In order to incorporate the skinning weight constraints, we draw inspiration from proximal algorithms \cite{parikh2014proximal}.
%experimented with softmax-type functions producing a probability distribution (ensuring non-negativity and partition of unity of the skinning weights) but there were two problems: 1) the convergence slows down when the logits get to nearly flat regions of the softmax; 2) we need another mechanism to ensure spatial sparsity, such as $L^1$ regularization, but this does not guarantee a fixed number of non-zeros among $w_{i,:}$. We tried to address these problems, but we found that a different approach, inspired by proximal algorithms \cite{parikh2014proximal}, worked best. 
The partition of unity constraints are easy: for each vertex $i$, we simply normalize the weights vector $w_{i,:}$ which ensures that $\sum_j w_{i,j} = 1$. The non-negativity and spatial sparsity constraints can be both satisfied by a single projection (proximal operator for an indicator function): for each weight vector $w_{i,:}$, we keep only the $K$ largest weights and zero out the negative ones. This is conveniently and efficiently implemented by a single call of the \texttt{torch.topk} function. This projection is performed after the Adam optimizer step and does not participate in auto-differentiation (in PyTorch this is accomplished via \texttt{torch.no\_grad}). 

With $p = 2$, \Eq{loss} is equivalent to the optimization problem solved by Dem Bones \cite{le2012smooth} and our approach converges to solutions with similar errors as the open source Dem Bones solver. \hl{The key advantage of our approach is the ease with which we can add additional constraints or change the loss function. 
%We can even reduce the number of proxy-bones to half and still obtain slightly better error than Dem Bones, but we are looking for more significant savings. 
We found that we can achieve significant savings by imposing sparsity constraints also on the $\B$ matrix (the sparsity of $\mat{C}$ is standard in all skinning decomposition methods).
We implement this additional sparsity constraint by another projection step, which consists in zeroing-out all but the largest $L$ elements of $\B$ in absolute value.} The absolute value is necessary because we need to allow negative values in the skinning transformations. Another difference to the $\mat{C}$-sparsity case is that with $\B$, \hl{we can distribute the non-zeros arbitrarily in the $\B$ matrix, whereas the $\mat{C}$ limits the non-zeros to $K$ per column, due to the limitations of GPUs and standard skinning pipelines.}
%can have a global budget for the non-zeros. This is because the implementation of \Eq{M} is our own and will typically run on the CPU, unlike linear blend skinning shaders that have to conform to existing skinning standards and limitations due to the GPUs and graphics pipelines.
However, the \texttt{torch.topk} function applied to \texttt{B.abs()} still works even in the case of the global (as opposed to per-column) budget of non-zero coefficients. In our experiments we typically set $L$ to 6000, which corresponds to 1000 transformations in the representation according to \Eq{rt}. This is less than 10\% of the transformations used by Dem Bones, but we are still able to get similar or even better accuracy (\Sec{Results}).

\subsection{High Detail (HD) Fit}

\hl{The flexibility of the PyTorch implementation invites experimentation with different values of $p$ in \mbox{\Eq{loss}}, corresponding to different norms. A particularly interesting case is}
%exploration of a secondary question: can we use our new optimization approach to significantly reduce the fitting error (\Eq{loss})? Ideally, we would use 
$p = \infty$, in which case \Eq{loss} minimizes the \textit{maximal} deviation from the ground truth blendshapes. We call this the ``HD'' (High Detail) setting because the infinity norm is more detail-sensitive.
%If our face models are of typical human sizes, we can argue that sufficiently small maximal error, e.g., less than $0.01$ centimeters, will be barely noticeable in the final rendered images. 
Naively using the infinity norm (max) actually works, but the convergence is extremely slow because only the vertex with the maximal error generates non-zero gradients. Instead, we can approximate the infinity norm with an $L^p$ norm with a high $p$; experimentally, we found that $p = 12$ works well if the model has enough capacity in terms of the number of non-zeros in $\B, \mat{C}$.
However, in our primary goal of reducing the compute overheads by maximizing the sparsity of $\B, \mat{C}$, we found that higher $p$ can produce non-smooth results; therefore, we use $p = 2$ by default. %However, we in applications where higher accuracy is required and more resources are available, using a larger $p$ may be advantageous (\Figure{proteus_high}).

%Therefore, in our primary task of reducing the memory footprint we recommend using the default $p = 2$. However, we thought that our findings with higher $p$ are worth reporting since they may be relevant in future work.

%Even though memory savings are our main objective, 

% - Error measurement, L_infinity / E_max
\section{Results}
\label{sec:Results}

\begin{figure}[h]
    \centering
    \includegraphics[trim={0 75mm 0 0},clip]{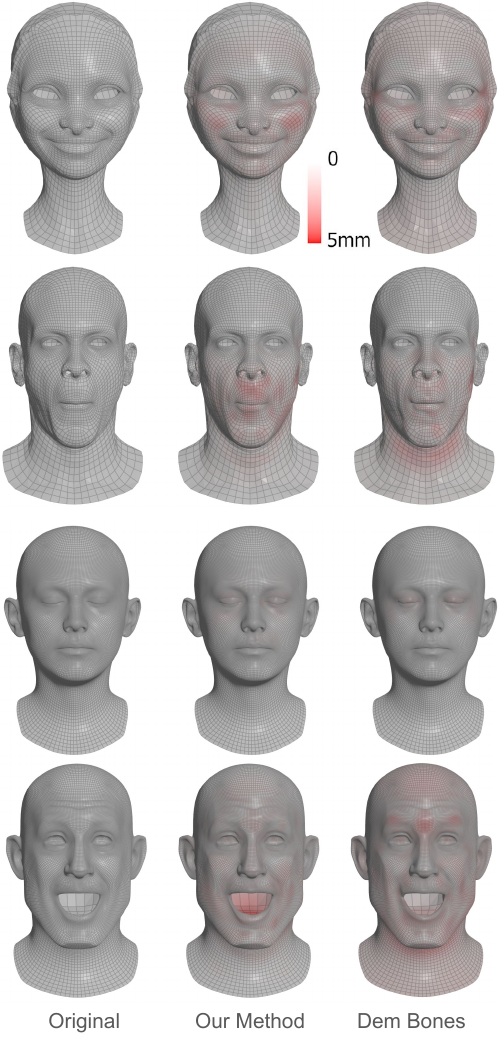}
    \caption{Our method leads to results of acceptable visual quality on various rigs and facial expressions, with errors comparable to Dem Bones (red color corresponds to error of 5mm or more). However, our method enables more efficient run-time.}
    \label{fig:results}
    %\vspace{-2mm}
\end{figure}

\begin{figure}[h]
    \centering
    \includegraphics[trim={0 192mm 0 0},clip]{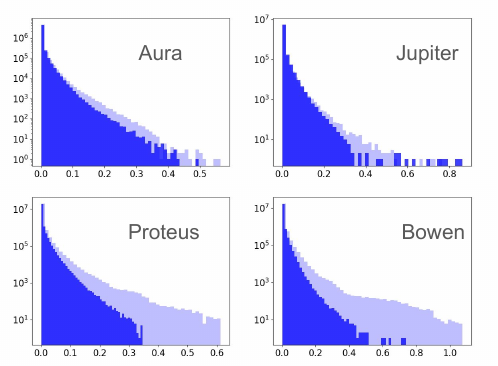}
    \caption{\hl{Histograms of the errors of our method (dark blue) and Dem Bones (light blue) in centimeters. Our method achieves lower errors despite sparse skining transformations.}}
    \label{fig:results_hists}
    %\vspace{-2mm}
\end{figure}

We use two error metrics to quantify the accuracy of a skinning decomposition; mean absolute error (MAE) and maximum absolute error (MXE).
The MAE is the mean of $E_{i,k}$ (\Eq{loss}) and tells us what error can we expect for a randomly chosen vertex and blendshape.
The MXE $= \max_{i,k} E_{i,k}$ measures the worst error. These metrics correspond to $L^1$ and $L^\infty$ norms.
Note that in our task, we assume that our input blendshapes have been carefully prepared and are thus treated as noise-free ground truth, i.e., we cannot dismiss large errors as outliers.

In our first set of experiments, we set the total number of proxy-bones to 40 for both Dem Bones and our method, but we limit our method to no more than 6000 non-zeros in the matrix $\B$.
This means that our method has about 10$\times$ fewer transformations to work with than Dem Bones, while achieving similar or better MAE and MXE (\Table{results}). We
set the number of influences $K$ to 8 for both Dem Bones and our method. Rendering of the resulting shapes confirms that the results of Dem Bones and our method are visually similar (\Figure{results}) and even harder to discern in an animation (see the accompanying video). \hl{We plot the corresponding error histograms in \mbox{\Figure{results_hists}}.}
%Red color corresponds to error 5mm or more, see the error bar in \Figure{results}.

\begin{table*}[htp]\centering
\caption{Statistics of our testing rigs and achieved fitting accuracy for our method and Dem Bones. Both the maximum (MXE) and mean (MAE) errors are in millimeters, with human-sized head models.}\label{tab:results}
%\scriptsize
\begin{tabular}{lrrrr D{.}{.}{7} D{.}{.}{7} r D{.}{.}{7} D{.}{.}{7}}\toprule
& & & &\multicolumn{3}{c}{Our method} &\multicolumn{3}{c}{Dem Bones} \\\cmidrule{5-10}
Model &Vertices ($N$) &Shapes ($S$) &Bones ($P$) &Transforms  &\multicolumn{1}{l}{MXE} &\multicolumn{1}{l}{MAE} &Transforms   &\multicolumn{1}{l}{MXE} &\multicolumn{1}{l}{MAE} \\\midrule
Aura &5944 &267 &40 &1000                           &5.82    &0.0384  &10680      &5.65   &0.0391 \\
Jupiter &5944 &319 &40 &1000                        &8.26    &0.0297  &12760      &7.64   &0.0263 \\
Proteus &23735 &287 &40 &1000                       &4.8     &0.03    &11480      &6.23   &0.0305 \\
Bowen &23735 &253 &40 &1000                           &5.99    &0.0339  &10120      &10.75  &0.0459 \\
%\hline
Proteus HD &23735 &287 &200 &57400                  &0.06    &0.0147  &57400      &3.45   &0.0174 \\
\bottomrule
\end{tabular}
\end{table*}

{To optimize our compressed skinning decompositions, we used 20k iterations of Adam with projection (\Sec{CSD}) with \hl{lr = $10^{-3}$, $\beta_1 = 0.9, \beta_2 = 0.9$}; the whole optimization takes several minutes on a single A6000 GPU. Dem Bones runs in under a minute on the CPU. The pre-processing times are relatively unimportant compared to the reduction of the run-time overhead of blending skinning transformations (\Eq{M}). This formula can be expressed as matrix multiplication of the coefficients $c_k$ with the matrix $\B$. With our method, the $\B$ is sparse, but with Dem Bones and other previous methods, the matrix $\B$ is dense.
Sparse matrix data structures (we use compressed row storage) need to store additional index information; we account for these in \Table{results_mem}. Despite this, sparse data structures offer $5$ - $7 \times$ memory savings. In \Table{results_speed}, we report the run-time performance measured on Snapdragon 652, which is a representative of our target low-spec mobile platforms. We can see that in addition to the memory savings, sparse storage offers also about $2$ - $3\times$ speed-up and brings the transformation blending times on par with rig evaluation (\Table{results_speed}). To put the timings in context, with 120Hz refresh rate, the total time budget for a frame is only about 8$ms$ and this needs to accommodate everything, including full-body animation, rendering, background and typically also multiple characters.

\begin{table}[htp]\centering
\caption{Memory requirements for sparse (our method) and dense transformations (Dem Bones), using 40 bones in both cases.}\label{tab:results_mem}
\begin{tabular}{lcccc}\toprule
Model   & Sparse            & Dense               & Ratio \\
\midrule
Aura    & 81k               & 512k    & 6.3$\times$  \\
Jupiter & 85k               & 612k    & 7.2$\times$  \\
Proteus & 85k               & 551k    & 6.5$\times$  \\
Bowen     & 87k               & 486k    & 5.6$\times$  \\
\bottomrule
\end{tabular}
\end{table}

\begin{table}[htp]\centering
\caption{Run-time speed measurements for sparse (our method) and dense transformations (Dem Bones), using 40 bones in both cases. We also report the rig evaluation time.}\label{tab:results_speed}
\begin{tabular}{lcccc}\toprule
Model   & Rig         & Sparse            & Dense               & Speed-up \\
\midrule
Aura    & 164$\mu s$  & 160$\mu s$  &  552$\mu s$   &  3.5$\times$ \\
Jupiter & 274$\mu s$  & 251$\mu s$  &  653$\mu s$   &  2.6$\times$ \\
Proteus & 201$\mu s$  & 171$\mu s$  &  585$\mu s$   &  3.4$\times$ \\
Bowen     & 159$\mu s$  & 185$\mu s$  &  520$\mu s$   &  2.8$\times$ \\
\bottomrule
\end{tabular}
\end{table}

Another option to consider instead of our method would be to decrease the number of bones $P$ in Dem Bones. We tried this with $P = 20, 10, 5$ and even $1$ (i.e., single transformation for the entire model, which we tried just out of curiosity, the visual quality is of course insufficient). The results are in \Table{results_DEM_vary}. We can see the errors increase rather quickly, e.g., with the Aura model, the MXE increased from 0.565 ($P = 40$) to 0.89 ($P = 20$) and the MAE increases similarly from 0.00391 to 0.0058 (\Figure{DEM_vary_bones}). Our method enables run-time efficiencies while achieving similar or even lower errors than Dem Bones. 

\hl{Another potential alternative to our method is to sparsify the skinning transformations after they have been computed by Dem Bones \mbox{\cite{dembones}}. To evaluate this approach, we have selected thresholds to zero-out translations and rotations to obtain similar sparsity as our method; specifically we have set the translation threshold as 1mm and the rotation threshold as 1 degree. These sparsified transformations lead to MAEs which are 1.5 to 3 times larger than our method (\mbox{\Table{results_sparse_DEM}}).} %However, the errors obtained with this approach are significantly higher because our method incorporates the sparsity constraints during optimization (as opposed to an after-the-fact sparsification).

\begin{figure}[h]
    \centering
    \includegraphics[trim={0cm 30mm 0 0},clip]{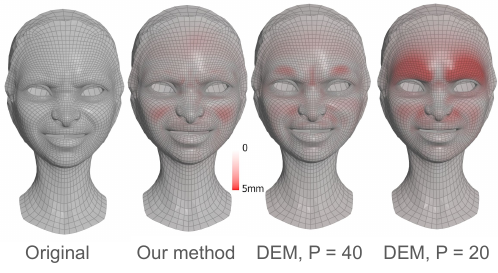}
    \caption{Decreasing the number of bones in Dem Bones from 40 to 20 increases the error significantly.}
    \label{fig:DEM_vary_bones}
\end{figure}

In our ``High-Detail'' experiment, we tried to minimize MXE by setting $p = 12$ in \Eq{loss}, using 200 proxy-bones, disabling the sparsification
of our transformations and setting $K$ to 32. We used the same settings in Dem Bones and set smoothness to zero to achieve the highest
accuracy. We used 500k iterations (2.5 hours on an A6000 GPU); Dem Bones still uses 
only about a minute on the CPU. The error metrics, reported as ``Proteus HD'' in \Table{results}, show that our $p = 12$ minimization achieved
more than 50$\times$ lower maximal error (MXE) than Dem Bones. The MAE errors are very similar ($0.00147$ vs. $0.00174$), suggesting that the maximal
error is well localized.
The visual impact of the larger MXE error is loss of details, such as the wrinkles caused by frowning (\Figure{proteus_high}).

\hl{We have also implemented our method in Unity and benchmarked the runtimes on a modern Windows PC (AMD 3975WX and NVIDIA RTX 3080). To stress-test the system, we are displaying 10 copies of each of our 4 characters (\mbox{\Figure{unity}}). Since both the CPU and GPU run concurrently, the final FPS is determined by the slower one; in all of our scenarios the CPU is the bottleneck (\mbox{\Table{results_PC}}). Our method is 1.4 times faster than Dem Bones as well as Unity's native implementation of blendshapes. This implementation is optimized for cache coherency etc., but requires 4.2 times longer GPU compute.}

\begin{figure}[h]
    \centering
    \includegraphics[trim={0cm 0mm 0 0},clip]{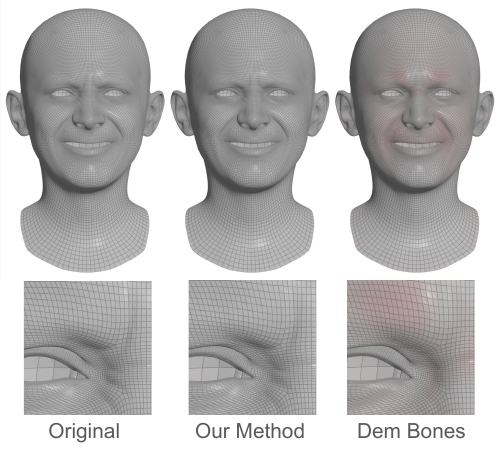}
    \caption{``Proteus HD'' experiment: our method with 57400 transforms and $L^{12}$ norm captures finer detail than Dem Bones with the same number of transformations.}
    \label{fig:proteus_high}
\end{figure}

\begin{figure}[h]
    \centering
    \includegraphics[scale=0.1,trim={0cm 0mm 0 0},clip]{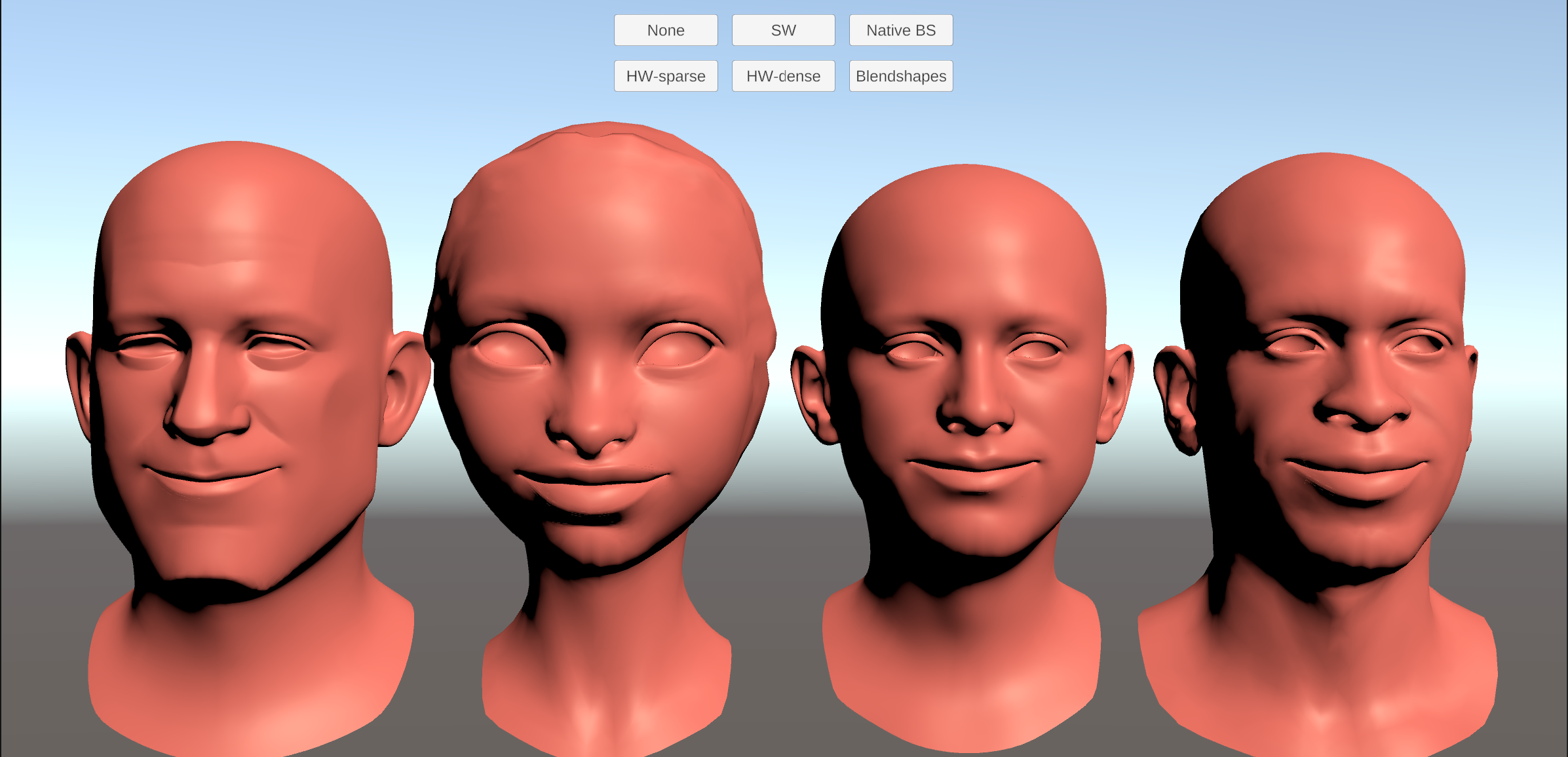}
    \caption{Screenshot of our Unity app.}
    \label{fig:unity}
\end{figure}

\begin{table}[htp]\centering
\caption{Dem Bones results with lower numbers of bones $P$ (MXE / MAE in millimeters).}\label{tab:results_DEM_vary}
\small
\begin{tabular}{lrrrr}\toprule
Model & $P = 20$ & $P = 10$ & $P = 5$ & $P = 1$ \\
\midrule
Aura & 8.9 / 0.058 & 21.0 / 0.096 & 21.2 / 0.113 & 33.7 / 0.225 \\
Jupiter & 10.7 / 0.045 & 11.6 / 0.063 & 11.8 / 0.077 & 31.2 / 0.155 \\
Proteus & 11.1 / 0.054 & 11.4 / 0.070 & 11.5 / 0.090 & 29.6 / 0.168 \\
Bowen & 9.1 / 0.053 & 12.9 / 0.089 & 15.0 / 0.106 & 33.5 / 0.215 \\
\bottomrule
\end{tabular}
\end{table}

\begin{table}[htp]\centering
\caption{\hl{Errors with sparsified Dem Bones transformations (40 bones) in millimeters.}}\label{tab:results_sparse_DEM}
\begin{tabular}{lc D{.}{.}{7} D{.}{.}{7}}\toprule
Model   & Non-zeros         & \multicolumn{1}{c}{MXE}     & \multicolumn{1}{c}{MAE} \\
\midrule
Aura    & 6767              & 7.46  & 0.12  \\
Jupiter & 5956              & 10.7  & 0.040  \\
Proteus & 6273              & 6.12  & 0.047  \\
Bowen     & 6520              & 10.72 & 0.12  \\
\bottomrule
\end{tabular}
\end{table}

\begin{table}[htp]\centering
\caption{\hl{Performance on a Windows PC in FPS and milliseconds.}}\label{tab:results_PC}
%\begin{tabular}{lc D{.}{.}{7} D{.}{.}{7}}\toprule
%Scenario            & FPS   & \multicolumn{1}{c}{CPU}     & \multicolumn{1}{c}{GPU} \\
\begin{tabular}{lccc}\toprule
Scenario            & FPS   & CPU   & GPU   \\
\midrule
Empty scene         & 2997  & 0.34  & 0.05  \\
Our method          & 252   & 3.97  & 1.08  \\
Dem Bones           & 180   & 5.35  & 1.18  \\
Unity blendshapes   & 185   & 5.41  & 4.61  \\
\bottomrule
\end{tabular}
\end{table}
\section{Conclusion}

We have presented a novel method for linear blend skinning decomposition and its integration into a facial animation pipeline.
Our new optimization strategy inspired by proximal algorithms outperforms the state-of-the-art and allows us to significantly
reduce the run-time overhead of blending skinning transformations, as well as the memory footprint thanks to sparse storage. Alternatively, the benefits of our new optimization method can also be
directed towards increasing the fitting accuracy and preserving geometric details. Our meshes contain eye- and mouth-bags and the error is evaluated equally
on all vertices. One possible extension of our method would be to introduce importance weighing of individual vertices to e.g. reduce the accuracy
on the usually invisible insides the eyes and the mouth and increase accuracy on more salient parts of the mesh, such as the nasolabial fold.

One limitation of our method are longer pre-processing times due to the first-order optimization approach that requires many iterations.
Another limitation is that we consider the rig function as a black box, even though there may be room for
further optimizations. In the future, it may be possible to jointly optimize the skinning decomposition along with
a neural network approximation of the rig function \cite{bailey2018fast,bailey2020fast,radzihovsky2020facebaker}; this would require 
us to adopt new rig evaluation mechanisms, but it could potentially unlock further efficiencies.

\begin{acks}

We thank Brian Budge, Roman Fedotov, Ryan Goldade, Stephane Grabli, Philipp Herholz, Petr Kadlecek, Binh Le, J.P. Lewis, Ronald Mallet, Olga Sorkine and Yuting Ye for inspiring discussions and the anonymous reviewers for constructive feedback.

\end{acks}

%\clearpage

% Bibliography
\bibliographystyle{ACM-Reference-Format}
\bibliography{bibliography}


\begin{thebibliography}{38}

%%% the Web address.

\ifx \showCODEN    \undefined \def \showCODEN     #1{\unskip}     \fi
\ifx \showDOI      \undefined \def \showDOI       #1{#1}\fi
\ifx \showISBNx    \undefined \def \showISBNx     #1{\unskip}     \fi
\ifx \showISBNxiii \undefined \def \showISBNxiii  #1{\unskip}     \fi
\ifx \showISSN     \undefined \def \showISSN      #1{\unskip}     \fi
\ifx \showLCCN     \undefined \def \showLCCN      #1{\unskip}     \fi
\ifx \shownote     \undefined \def \shownote      #1{#1}          \fi
\ifx \showarticletitle \undefined \def \showarticletitle #1{#1}   \fi
\ifx \showURL      \undefined \def \showURL       {\relax}        \fi
% invisible to TeX
\providecommand\bibfield[2]{#2}
\providecommand\bibinfo[2]{#2}
\providecommand\natexlab[1]{#1}
\providecommand\showeprint[2][]{arXiv:#2}

\bibitem[Aharon et~al\mbox{.}(2006)]%
        {aharon2006}
\bibfield{author}{\bibinfo{person}{Michal Aharon}, \bibinfo{person}{Michael Elad}, {and} \bibinfo{person}{Alfred Bruckstein}.} \bibinfo{year}{2006}\natexlab{}.
\newblock \showarticletitle{K-SVD: An algorithm for designing overcomplete dictionaries for sparse representation}.
\newblock \bibinfo{journal}{\emph{IEEE Transactions on signal processing}} \bibinfo{volume}{54}, \bibinfo{number}{11} (\bibinfo{year}{2006}), \bibinfo{pages}{4311--4322}.
\newblock


\bibitem[Bailey et~al\mbox{.}(2020)]%
        {bailey2020fast}
\bibfield{author}{\bibinfo{person}{Stephen~W. Bailey}, \bibinfo{person}{Dalton Omens}, \bibinfo{person}{Paul Dilorenzo}, {and} \bibinfo{person}{James~F. O'Brien}.} \bibinfo{year}{2020}\natexlab{}.
\newblock \showarticletitle{Fast and Deep Facial Deformations}.
\newblock \bibinfo{journal}{\emph{ACM Transactions on Graphics}} \bibinfo{volume}{39}, \bibinfo{number}{4} (\bibinfo{date}{Aug.} \bibinfo{year}{2020}), \bibinfo{pages}{94:1--15}.
\newblock
\urldef\tempurl%
\url{https://doi.org/10.1145/3386569.3392397}
\showDOI{\tempurl}
\newblock
\shownote{Presented at SIGGRAPH 2020, Washington D.C.}.


\bibitem[Bailey et~al\mbox{.}(2018)]%
        {bailey2018fast}
\bibfield{author}{\bibinfo{person}{Stephen~W Bailey}, \bibinfo{person}{Dave Otte}, \bibinfo{person}{Paul Dilorenzo}, {and} \bibinfo{person}{James~F O'Brien}.} \bibinfo{year}{2018}\natexlab{}.
\newblock \showarticletitle{Fast and deep deformation approximations}.
\newblock \bibinfo{journal}{\emph{ACM Transactions on Graphics (TOG)}} \bibinfo{volume}{37}, \bibinfo{number}{4} (\bibinfo{year}{2018}), \bibinfo{pages}{1--12}.
\newblock


\bibitem[Bharaj et~al\mbox{.}(2012)]%
        {bharaj2012automatically}
\bibfield{author}{\bibinfo{person}{Gaurav Bharaj}, \bibinfo{person}{Thorsten Thorm{\"a}hlen}, \bibinfo{person}{Hans-Peter Seidel}, {and} \bibinfo{person}{Christian Theobalt}.} \bibinfo{year}{2012}\natexlab{}.
\newblock \showarticletitle{Automatically rigging multi-component characters}. In \bibinfo{booktitle}{\emph{Computer Graphics Forum}}, Vol.~\bibinfo{volume}{31}. \bibinfo{pages}{755--764}.
\newblock


\bibitem[Buss and Fillmore(2001)]%
        {buss2001spherical}
\bibfield{author}{\bibinfo{person}{Samuel~R Buss} {and} \bibinfo{person}{Jay~P Fillmore}.} \bibinfo{year}{2001}\natexlab{}.
\newblock \showarticletitle{Spherical averages and applications to spherical splines and interpolation}.
\newblock \bibinfo{journal}{\emph{ACM Transactions on Graphics (TOG)}} \bibinfo{volume}{20}, \bibinfo{number}{2} (\bibinfo{year}{2001}), \bibinfo{pages}{95--126}.
\newblock


\bibitem[Chandran et~al\mbox{.}(2022)]%
        {chandran2022facial}
\bibfield{author}{\bibinfo{person}{Prashanth Chandran}, \bibinfo{person}{Gaspard Zoss}, \bibinfo{person}{Markus Gross}, \bibinfo{person}{Paulo Gotardo}, {and} \bibinfo{person}{Derek Bradley}.} \bibinfo{year}{2022}\natexlab{}.
\newblock \showarticletitle{Facial Animation with Disentangled Identity and Motion using Transformers}. In \bibinfo{booktitle}{\emph{Computer Graphics Forum}}, Vol.~\bibinfo{volume}{41}. Wiley Online Library, \bibinfo{pages}{267--277}.
\newblock


\bibitem[Choi et~al\mbox{.}(2022)]%
        {choi2022animatomy}
\bibfield{author}{\bibinfo{person}{Byungkuk Choi}, \bibinfo{person}{Haekwang Eom}, \bibinfo{person}{Benjamin Mouscadet}, \bibinfo{person}{Stephen Cullingford}, \bibinfo{person}{Kurt Ma}, \bibinfo{person}{Stefanie Gassel}, \bibinfo{person}{Suzi Kim}, \bibinfo{person}{Andrew Moffat}, \bibinfo{person}{Millicent Maier}, \bibinfo{person}{Marco Revelant}, {et~al\mbox{.}}} \bibinfo{year}{2022}\natexlab{}.
\newblock \showarticletitle{Animatomy: An animator-centric, anatomically inspired system for 3d facial modeling, animation and transfer}. In \bibinfo{booktitle}{\emph{SIGGRAPH Asia 2022 Conference Papers}}. \bibinfo{pages}{1--9}.
\newblock


\bibitem[Costigan et~al\mbox{.}(2016)]%
        {costigan2016improving}
\bibfield{author}{\bibinfo{person}{Timothy Costigan}, \bibinfo{person}{Anton Gerdelan}, \bibinfo{person}{Emma Carrigan}, {and} \bibinfo{person}{Rachel McDonnell}.} \bibinfo{year}{2016}\natexlab{}.
\newblock \showarticletitle{Improving blendshape performance for crowds with GPU and GPGPU techniques}. In \bibinfo{booktitle}{\emph{Proceedings of the 9th International Conference on Motion in Games}}. \bibinfo{pages}{73--78}.
\newblock


\bibitem[Donoho(2006)]%
        {donoho2006compressed}
\bibfield{author}{\bibinfo{person}{David~L Donoho}.} \bibinfo{year}{2006}\natexlab{}.
\newblock \showarticletitle{Compressed sensing}.
\newblock \bibinfo{journal}{\emph{IEEE Transactions on information theory}} \bibinfo{volume}{52}, \bibinfo{number}{4} (\bibinfo{year}{2006}), \bibinfo{pages}{1289--1306}.
\newblock


\bibitem[Ekman and Friesen(1978)]%
        {ekman1978facial}
\bibfield{author}{\bibinfo{person}{Paul Ekman} {and} \bibinfo{person}{Wallace~V Friesen}.} \bibinfo{year}{1978}\natexlab{}.
\newblock \showarticletitle{Facial action coding system}.
\newblock \bibinfo{journal}{\emph{Environmental Psychology \& Nonverbal Behavior}} (\bibinfo{year}{1978}).
\newblock


\bibitem[{Electronic Arts}({[n.\,d.]})]%
        {dembones}
\bibfield{author}{\bibinfo{person}{{Electronic Arts}}.} \bibinfo{year}{[n.\,d.]}\natexlab{}.
\newblock \bibinfo{booktitle}{\emph{Dem Bones: an Open Source Library for Skinning Decomposition}}.
\newblock
\urldef\tempurl%
\url{https://www.ea.com/seed/news/open-source-dem-bones}
\showURL{%
\tempurl}


\bibitem[Goldstein et~al\mbox{.}(2002)]%
        {goldstein2002classical}
\bibfield{author}{\bibinfo{person}{H. Goldstein}, \bibinfo{person}{C.P. Poole}, {and} \bibinfo{person}{J.L. Safko}.} \bibinfo{year}{2002}\natexlab{}.
\newblock \bibinfo{booktitle}{\emph{Classical Mechanics}}.
\newblock \bibinfo{publisher}{Addison Wesley}.
\newblock
\showISBNx{9780201657029}
\showLCCN{79023456}
\urldef\tempurl%
\url{https://books.google.ch/books?id=tJCuQgAACAAJ}
\showURL{%
\tempurl}


\bibitem[Hasler et~al\mbox{.}(2010)]%
        {hasler2010learning}
\bibfield{author}{\bibinfo{person}{Nils Hasler}, \bibinfo{person}{Thorsten Thorm{\"a}hlen}, \bibinfo{person}{Bodo Rosenhahn}, {and} \bibinfo{person}{Hans-Peter Seidel}.} \bibinfo{year}{2010}\natexlab{}.
\newblock \showarticletitle{Learning skeletons for shape and pose}. In \bibinfo{booktitle}{\emph{Proceedings of the 2010 ACM SIGGRAPH symposium on Interactive 3D Graphics and Games}}. \bibinfo{pages}{23--30}.
\newblock


\bibitem[Hoyer(2004)]%
        {hoyer2004non}
\bibfield{author}{\bibinfo{person}{Patrik~O Hoyer}.} \bibinfo{year}{2004}\natexlab{}.
\newblock \showarticletitle{Non-negative matrix factorization with sparseness constraints.}
\newblock \bibinfo{journal}{\emph{Journal of machine learning research}} \bibinfo{volume}{5}, \bibinfo{number}{9} (\bibinfo{year}{2004}).
\newblock


\bibitem[Jacobson et~al\mbox{.}(2011)]%
        {jacobson2011bounded}
\bibfield{author}{\bibinfo{person}{Alec Jacobson}, \bibinfo{person}{Ilya Baran}, \bibinfo{person}{Jovan Popovic}, {and} \bibinfo{person}{Olga Sorkine}.} \bibinfo{year}{2011}\natexlab{}.
\newblock \showarticletitle{Bounded biharmonic weights for real-time deformation.}
\newblock \bibinfo{journal}{\emph{ACM Trans. Graph.}} \bibinfo{volume}{30}, \bibinfo{number}{4} (\bibinfo{year}{2011}), \bibinfo{pages}{78}.
\newblock


\bibitem[James and Twigg(2005)]%
        {james2005skinning}
\bibfield{author}{\bibinfo{person}{Doug~L James} {and} \bibinfo{person}{Christopher~D Twigg}.} \bibinfo{year}{2005}\natexlab{}.
\newblock \showarticletitle{Skinning mesh animations}.
\newblock \bibinfo{journal}{\emph{ACM Transactions on Graphics (TOG)}} \bibinfo{volume}{24}, \bibinfo{number}{3} (\bibinfo{year}{2005}), \bibinfo{pages}{399--407}.
\newblock


\bibitem[Kavan et~al\mbox{.}(2007)]%
        {kavan2007skinning}
\bibfield{author}{\bibinfo{person}{Ladislav Kavan}, \bibinfo{person}{Rachel McDonnell}, \bibinfo{person}{Simon Dobbyn}, \bibinfo{person}{Ji{\v{r}}{\'\i} {\v{Z}}{\'a}ra}, {and} \bibinfo{person}{Carol O'Sullivan}.} \bibinfo{year}{2007}\natexlab{}.
\newblock \showarticletitle{Skinning arbitrary deformations}. In \bibinfo{booktitle}{\emph{Proceedings of the 2007 symposium on Interactive 3D graphics and games}}. \bibinfo{pages}{53--60}.
\newblock


\bibitem[Kavan et~al\mbox{.}(2010)]%
        {kavan2010fast}
\bibfield{author}{\bibinfo{person}{Ladislav Kavan}, \bibinfo{person}{P-P Sloan}, {and} \bibinfo{person}{Carol O'Sullivan}.} \bibinfo{year}{2010}\natexlab{}.
\newblock \showarticletitle{Fast and efficient skinning of animated meshes}. In \bibinfo{booktitle}{\emph{Computer Graphics Forum}}, Vol.~\bibinfo{volume}{29}. Wiley Online Library, \bibinfo{pages}{327--336}.
\newblock


\bibitem[Kim and Singh(2021)]%
        {kim2021optimizing}
\bibfield{author}{\bibinfo{person}{Joonho Kim} {and} \bibinfo{person}{Karan Singh}.} \bibinfo{year}{2021}\natexlab{}.
\newblock \showarticletitle{Optimizing UI layouts for deformable face-rig manipulation}.
\newblock \bibinfo{journal}{\emph{ACM Transactions on Graphics (TOG)}} \bibinfo{volume}{40}, \bibinfo{number}{4} (\bibinfo{year}{2021}), \bibinfo{pages}{1--12}.
\newblock


\bibitem[Kingma and Ba(2014)]%
        {kingma2014adam}
\bibfield{author}{\bibinfo{person}{Diederik~P Kingma} {and} \bibinfo{person}{Jimmy Ba}.} \bibinfo{year}{2014}\natexlab{}.
\newblock \showarticletitle{Adam: A method for stochastic optimization}.
\newblock \bibinfo{journal}{\emph{arXiv preprint arXiv:1412.6980}} (\bibinfo{year}{2014}).
\newblock


\bibitem[Le and Deng(2012)]%
        {le2012smooth}
\bibfield{author}{\bibinfo{person}{Binh~Huy Le} {and} \bibinfo{person}{Zhigang Deng}.} \bibinfo{year}{2012}\natexlab{}.
\newblock \showarticletitle{Smooth Skinning Decomposition with Rigid Bones}.
\newblock \bibinfo{journal}{\emph{ACM Trans. Graph.}} \bibinfo{volume}{31}, \bibinfo{number}{6} (\bibinfo{year}{2012}).
\newblock


\bibitem[Le and Deng(2013)]%
        {le2013two}
\bibfield{author}{\bibinfo{person}{Binh~Huy Le} {and} \bibinfo{person}{Zhigang Deng}.} \bibinfo{year}{2013}\natexlab{}.
\newblock \showarticletitle{Two-layer sparse compression of dense-weight blend skinning}.
\newblock \bibinfo{journal}{\emph{ACM Transactions on Graphics (TOG)}} \bibinfo{volume}{32}, \bibinfo{number}{4} (\bibinfo{year}{2013}), \bibinfo{pages}{1--10}.
\newblock


\bibitem[Le and Deng(2014)]%
        {le2014robust}
\bibfield{author}{\bibinfo{person}{Binh~Huy Le} {and} \bibinfo{person}{Zhigang Deng}.} \bibinfo{year}{2014}\natexlab{}.
\newblock \showarticletitle{Robust and accurate skeletal rigging from mesh sequences}.
\newblock \bibinfo{journal}{\emph{ACM Transactions on Graphics (TOG)}} \bibinfo{volume}{33}, \bibinfo{number}{4} (\bibinfo{year}{2014}), \bibinfo{pages}{1--10}.
\newblock


\bibitem[Lee and Seung(1999)]%
        {lee1999learning}
\bibfield{author}{\bibinfo{person}{Daniel~D Lee} {and} \bibinfo{person}{H~Sebastian Seung}.} \bibinfo{year}{1999}\natexlab{}.
\newblock \showarticletitle{Learning the parts of objects by non-negative matrix factorization}.
\newblock \bibinfo{journal}{\emph{Nature}} \bibinfo{volume}{401}, \bibinfo{number}{6755} (\bibinfo{year}{1999}), \bibinfo{pages}{788--791}.
\newblock


\bibitem[Lewis et~al\mbox{.}(2014)]%
        {lewis2014practice}
\bibfield{author}{\bibinfo{person}{John~P Lewis}, \bibinfo{person}{Ken Anjyo}, \bibinfo{person}{Taehyun Rhee}, \bibinfo{person}{Mengjie Zhang}, \bibinfo{person}{Frederic~H Pighin}, {and} \bibinfo{person}{Zhigang Deng}.} \bibinfo{year}{2014}\natexlab{}.
\newblock \showarticletitle{Practice and theory of blendshape facial models.}
\newblock \bibinfo{journal}{\emph{Eurographics (State of the Art Reports)}} \bibinfo{volume}{1}, \bibinfo{number}{8} (\bibinfo{year}{2014}), \bibinfo{pages}{2}.
\newblock


\bibitem[Li et~al\mbox{.}(2017)]%
        {tianye2017flame}
\bibfield{author}{\bibinfo{person}{Tianye Li}, \bibinfo{person}{Timo Bolkart}, \bibinfo{person}{Michael.~J. Black}, \bibinfo{person}{Hao Li}, {and} \bibinfo{person}{Javier Romero}.} \bibinfo{year}{2017}\natexlab{}.
\newblock \showarticletitle{Learning a model of facial shape and expression from {4D} scans}.
\newblock \bibinfo{journal}{\emph{ACM Transactions on Graphics, (Proc. SIGGRAPH Asia)}} \bibinfo{volume}{36}, \bibinfo{number}{6} (\bibinfo{year}{2017}), \bibinfo{pages}{194:1--194:17}.
\newblock
\urldef\tempurl%
\url{https://doi.org/10.1145/3130800.3130813}
\showURL{%
\tempurl}


\bibitem[Magnenat-Thalmann et~al\mbox{.}(1988)]%
        {magnenat1988joint}
\bibfield{author}{\bibinfo{person}{Nadia Magnenat-Thalmann}, \bibinfo{person}{Richard Laperrire}, {and} \bibinfo{person}{Daniel Thalmann}.} \bibinfo{year}{1988}\natexlab{}.
\newblock \showarticletitle{Joint-dependent local deformations for hand animation and object grasping}. In \bibinfo{booktitle}{\emph{In Proceedings on Graphics Interface 1988}}.
\newblock


\bibitem[Melchior et~al\mbox{.}(2019)]%
        {melchior2019proximal}
\bibfield{author}{\bibinfo{person}{Peter Melchior}, \bibinfo{person}{R{\'e}my Joseph}, {and} \bibinfo{person}{Fred Moolekamp}.} \bibinfo{year}{2019}\natexlab{}.
\newblock \showarticletitle{Proximal Adam: robust adaptive update scheme for constrained optimization}.
\newblock \bibinfo{journal}{\emph{arXiv preprint arXiv:1910.10094}} (\bibinfo{year}{2019}).
\newblock


\bibitem[Meyer and Anderson(2007)]%
        {meyer2007key}
\bibfield{author}{\bibinfo{person}{Mark Meyer} {and} \bibinfo{person}{John Anderson}.} \bibinfo{year}{2007}\natexlab{}.
\newblock \showarticletitle{Key point subspace acceleration and soft caching}.
\newblock In \bibinfo{booktitle}{\emph{ACM SIGGRAPH 2007 papers}}. \bibinfo{pages}{74--es}.
\newblock


\bibitem[Moutafidou et~al\mbox{.}(2023)]%
        {moutafidou2023deep}
\bibfield{author}{\bibinfo{person}{Anastasia Moutafidou}, \bibinfo{person}{Vasileios Toulatzis}, {and} \bibinfo{person}{Ioannis Fudos}.} \bibinfo{year}{2023}\natexlab{}.
\newblock \showarticletitle{Deep fusible skinning of animation sequences}.
\newblock \bibinfo{journal}{\emph{The Visual Computer}} (\bibinfo{year}{2023}), \bibinfo{pages}{1--21}.
\newblock


\bibitem[Neumann et~al\mbox{.}(2013)]%
        {neumann2013sparse}
\bibfield{author}{\bibinfo{person}{Thomas Neumann}, \bibinfo{person}{Kiran Varanasi}, \bibinfo{person}{Stephan Wenger}, \bibinfo{person}{Markus Wacker}, \bibinfo{person}{Marcus Magnor}, {and} \bibinfo{person}{Christian Theobalt}.} \bibinfo{year}{2013}\natexlab{}.
\newblock \showarticletitle{Sparse localized deformation components}.
\newblock \bibinfo{journal}{\emph{ACM Transactions on Graphics (TOG)}} \bibinfo{volume}{32}, \bibinfo{number}{6} (\bibinfo{year}{2013}), \bibinfo{pages}{1--10}.
\newblock


\bibitem[Parikh et~al\mbox{.}(2014)]%
        {parikh2014proximal}
\bibfield{author}{\bibinfo{person}{Neal Parikh}, \bibinfo{person}{Stephen Boyd}, {et~al\mbox{.}}} \bibinfo{year}{2014}\natexlab{}.
\newblock \showarticletitle{Proximal algorithms}.
\newblock \bibinfo{journal}{\emph{Foundations and trends{\textregistered} in Optimization}} \bibinfo{volume}{1}, \bibinfo{number}{3} (\bibinfo{year}{2014}), \bibinfo{pages}{127--239}.
\newblock


\bibitem[Paszke et~al\mbox{.}(2019)]%
        {paszke2019pytorch}
\bibfield{author}{\bibinfo{person}{Adam Paszke}, \bibinfo{person}{Sam Gross}, \bibinfo{person}{Francisco Massa}, \bibinfo{person}{Adam Lerer}, \bibinfo{person}{James Bradbury}, \bibinfo{person}{Gregory Chanan}, \bibinfo{person}{Trevor Killeen}, \bibinfo{person}{Zeming Lin}, \bibinfo{person}{Natalia Gimelshein}, \bibinfo{person}{Luca Antiga}, {et~al\mbox{.}}} \bibinfo{year}{2019}\natexlab{}.
\newblock \showarticletitle{Pytorch: An imperative style, high-performance deep learning library}.
\newblock \bibinfo{journal}{\emph{Advances in neural information processing systems}}  \bibinfo{volume}{32} (\bibinfo{year}{2019}).
\newblock


\bibitem[Radzihovsky et~al\mbox{.}(2020)]%
        {radzihovsky2020facebaker}
\bibfield{author}{\bibinfo{person}{Sarah Radzihovsky}, \bibinfo{person}{Fernando de Goes}, {and} \bibinfo{person}{Mark Meyer}.} \bibinfo{year}{2020}\natexlab{}.
\newblock \showarticletitle{Facebaker: Baking character facial rigs with machine learning}. In \bibinfo{booktitle}{\emph{Special Interest Group on Computer Graphics and Interactive Techniques Conference Talks}}. \bibinfo{pages}{1--2}.
\newblock


\bibitem[Schaefer and Yuksel(2007)]%
        {schaefer2007example}
\bibfield{author}{\bibinfo{person}{Scott Schaefer} {and} \bibinfo{person}{Can Yuksel}.} \bibinfo{year}{2007}\natexlab{}.
\newblock \showarticletitle{Example-based skeleton extraction}. In \bibinfo{booktitle}{\emph{Symposium on Geometry Processing}}. \bibinfo{pages}{153--162}.
\newblock


\bibitem[Seo et~al\mbox{.}(2011)]%
        {seo2011compression}
\bibfield{author}{\bibinfo{person}{Jaewoo Seo}, \bibinfo{person}{Geoffrey Irving}, \bibinfo{person}{John~P Lewis}, {and} \bibinfo{person}{Junyong Noh}.} \bibinfo{year}{2011}\natexlab{}.
\newblock \showarticletitle{Compression and direct manipulation of complex blendshape models}.
\newblock \bibinfo{journal}{\emph{ACM Transactions on Graphics (TOG)}} \bibinfo{volume}{30}, \bibinfo{number}{6} (\bibinfo{year}{2011}), \bibinfo{pages}{1--10}.
\newblock


\bibitem[Tarini et~al\mbox{.}(2014)]%
        {tarini2014accurate}
\bibfield{author}{\bibinfo{person}{Marco Tarini}, \bibinfo{person}{Daniele Panozzo}, {and} \bibinfo{person}{Olga Sorkine-Hornung}.} \bibinfo{year}{2014}\natexlab{}.
\newblock \showarticletitle{Accurate and efficient lighting for skinned models}. In \bibinfo{booktitle}{\emph{Computer Graphics Forum}}, Vol.~\bibinfo{volume}{33}. Wiley Online Library, \bibinfo{pages}{421--428}.
\newblock


\bibitem[Tibshirani(1996)]%
        {tibshirani1996regression}
\bibfield{author}{\bibinfo{person}{Robert Tibshirani}.} \bibinfo{year}{1996}\natexlab{}.
\newblock \showarticletitle{Regression shrinkage and selection via the lasso}.
\newblock \bibinfo{journal}{\emph{Journal of the Royal Statistical Society Series B: Statistical Methodology}} \bibinfo{volume}{58}, \bibinfo{number}{1} (\bibinfo{year}{1996}), \bibinfo{pages}{267--288}.
\newblock


\end{thebibliography}

\end{document}